# $Bi_2Te_{1.6}S_{1.4}$ – a Topological Insulator in the Tetradymite Family

Huiwen Ji[1], J.M. Allred[1], M.K. Fuccillo[1], M.E. Charles[1], M. Neupane[2], L.A. Wray[2,3], M.Z. Hasan[2] and R.J. Cava[1]

[1]Department of Chemistry, Princeton University, Princeton, NJ 08544
[2]Department of Physics, Princeton University, Princeton, NJ 08544
[3]Advanced Light Source, Lawrence Berkeley National Laboratory, Berkeley, CA 94305

**Abstract**

We describe the crystal growth, crystal structure, and basic electrical properties of $Bi_2Te_{1.6}S_{1.4}$, which incorporates both S and Te in its Tetradymite quintuple layers in the motif -$[Te_{0.8}S_{0.2}]$-Bi-S-Bi-$[Te_{0.8}S_{0.2}]$-. This material differs from other Tetradymites studied as topological insulators due to the increased ionic character that arises from its significant S content. $Bi_2Te_{1.6}S_{1.4}$ forms high quality crystals from the melt and is the S-rich limit of the ternary Bi-Te-S γ-Tetradymite phase at the melting point. The native material is *n*-type with a low resistivity; Sb substitution, with adjustment of the Te to S ratio, results in a crossover to *p*-type and resistive behavior at low temperatures. Angle resolved photoemission study shows that topological surface states are present, with the Dirac point more exposed than it is in $Bi_2Te_3$ and similar to that seen in $Bi_2Te_2Se$. Single crystal structure determination indicates that the S in the outer chalcogen layers is closer to the Bi than the Te, and therefore that the layers supporting the surface states are corrugated on the atomic scale.

**Introduction**

Topological Insulators (TIs), bulk insulators hosting time-reversal-protected spin-polarized surface states, are currently of significant interest.[1-3] These materials exhibit exotic physics[4-8] and may have prospects for applications in spintronics and quantum computing.[9] Thus far, however, relatively few three dimensional (3D) TIs have been experimentally characterized. These include binary compounds such as $Bi_2Se_3$ and $Bi_2Te_3$, the Bi-Sb alloy, and the ternary compound $TlBiSe_2$.[10-14] Several theoretical groups have pointed out that ternary variants of the $Bi_2Ch_3$ (Ch=chalcogen) Tetradymite-type compounds,[15, 16] such as $Bi_2Te_2Se$ and $Bi_2Te_2S$, should also be TIs. $Bi_2Te_2Se$ and its doped variants have in fact proven to be the most resistive 3D TIs to date.[17-19] The $Bi_2Te_2S$ Tetradymite, has not, however, been studied experimentally from the TI perspective, and the current lack of an ideal material for TI surface state transport studies is motivation for continuing to look for new materials. This phase is of particular interest because the substitution of more electronegative S for Se may result in lower bulk valence band energies than for $Bi_2Te_2Se$, potentially resulting in the exposure of the Dirac point in the surface states to a position above the top of the bulk valence band.

The formula of Tetradymite was first determined as "$Bi_2Te_2S$" from minerals in 1934,[20] although arguments eventually emerged about whether the formula made chemical sense[21, 22].[21, 22] Here we report the synthesis, crystal growth and characterization of a TI based on this compound, with a formula of $Bi_2Te_{1.6}S_{1.4}$, known as the γ phase-tetradymite.[22] It is composed of quintuple-layers of the type [Te/S]-Bi-S-Bi-[Te/S] and forms a compound distinct from either $Bi_2Te_3$ or $Bi_2S_3$. The high quality crystals of γ Tetradymite grow at a higher S content, $Bi_2Te_{1.6}S_{1.4}$, than is stable for materials annealed in the solid state, which have the composition $Bi_2Te_{1.89}S_{1.11}$. $Bi_2Te_{1.6}S_{1.4}$ is determined to be *n*-type by Hall measurements. Partial Sb substitution for Bi[23] leads to *p*-type behavior and a resistivity at 10 K of 0.1 Ohm-cm for the composition $Bi_{1.1}Sb_{0.9}Te_2S$. A single surface state Dirac cone is observed at the Γ point on the basal plane surface of single crystals of $Bi_2Te_{1.6}S_{1.4}$ by ARPES (Angle-Resolved Photoemission Spectroscopy), indicating that, as for other members of the family, it is a TI. The ARPES data also shows that the Dirac point is more exposed relative to the bulk valence band than in $Bi_2Te_3$ and is similar to what is seen in $Bi_2Te_2Se$.

**Experimental**

High purity elemental Bi (99.999%), Te (99.9999%), and S (99.999%) were used for the $Bi_2Te_{2-x}S_{1+x}$ crystal growth, initially with $x$ in the range of 0 to 1. Three-gram mixtures of pre-cleaned elements were sealed in clean quartz ampoules, heated up to 850 °C for 1 day followed by cooling over the period of an hour to 700 °C. The crystal growth for $Bi_2Te_{2-x}S_{1+x}$ involved cooling from 700 °C to 550 °C over a period of 25 h, followed by water quenching. High quality, large single crystals with high luster on their cleavage faces were obtained only for $x$ of 0.5. For comparison purposes, some crystals with $x = 0.5$ were annealed afterwards for 3 days in the solid state at 500 °C. As part of the doping study to see whether $p$-type behavior could be obtained, $Bi_{1.4}Sb_{0.6}Te_{1.8}S_{1.2}$ crystals were grown in the same fashion, from a melt with the 1.2:1.8 S to Te ratio, and $Bi_{1.1}Sb_{0.9}Te_2S$ crystals were grown from a melt of 1:2 S to Te ratio. For these substitutions, high purity Sb (99.999%) was employed.

Powder X-ray diffraction was performed on a Bruker D8 Focus X-ray diffractometer operating with Cu Kα radiation and a graphite diffracted beam monochromator. Single crystal X-ray diffraction data (SXRD) was collected on a Bruker APEX II using Mo Kα radiation (λ = 0.71073 Å) at room temperature. Unit cell determination and refinement, and data integration were performed with Bruker APEX2 software. The crystal structure was determined using SHELXL-97 implemented through WinGX. Specimens for single crystal work were obtained by breaking pieces off the boule under liquid nitrogen in order to minimize the effect of strain on the crystallinity.

Resistivity and Hall measurements were performed in a Quantum Design Physical Property Measurement System (PPMS). High-resolution angle-resolved photoemission spectroscopy (ARPES) measurements were performed at 20 K using 8-22 eV photon energies on beam line 5 at the Stanford Synchrotron Radiation Laboratory, California. The energy and momentum resolutions were 15 meV and 2% of the surface Brillouin zone, respectively, obtained using a Scienta R4000 analyzer. The samples were cleaved at 20 K under pressures of less than $5 \times 10^{-11}$ torr, resulting in shiny flat surfaces. For one sample, $Bi_{1.1}Sb_{0.9}Te_2Se$, 60 eV photon energies were employed to collect the ARPES spectra.

**Results**

Full structural characterization using single crystal diffraction was employed on crystals grown from a melt with a nominal composition $Bi_2Te_{1.5}S_{1.5}$. The experimental and refinement details are given in Table 1(a). The composition of the crystals, easily determined in the refinements due to the difference in scattering factors for Te and S, was found to be $Bi_2Te_{1.584}S_{1.416}$, with the uncertainty in S content of ±0.016 per formula unit. Therefore, we designate the formula as $Bi_2Te_{1.6}S_{1.4}$. $Bi_2Te_{1.6}S_{1.4}$ has the same space group symmetry, R-3m, as $Bi_2Se_3$ and $Bi_2Te_3$,[24] and like them the crystal structure consists of Ch(2)-Bi-Ch(1)-Bi-Ch(2) quintuple-layers with weak van der Waals bonding between them.

The refined atomic positions and occupancies for the final structural model for the crystal structure of $Bi_2Te_{1.6}S_{1.4}$ are given in Table 1(b). Consistent with earlier determinations, the inner chalcogen layer Ch(1), which is bonded to two sandwiching Bi layers, is solely occupied by S atoms, while the outer chalcogen layers (Ch(2)) are a mixture of S atoms and Te atoms. Careful structural refinement reveals that the S atoms and Te atoms in the Ch(2) layers are not in the same plane. The 80%Te-20%S outer layers are randomly corrugated at the atomic scale: as expected from simple ionic size considerations, the S atoms are closer to the Bi atoms than the Te atoms are. The superiority of this structural model is evidenced by comparing the refinement agreement (R) values as well as the thermal parameters for two structural models, one of which has both Te and S at the same position (table 1(c)) and the other allowing their positions to be different and freely refined (table 1(b)). The comparison shows that the model with the independent S(2)/Te(2) positions has significantly better agreement values ($R_1$=0.0251, $wR_2$=0.0700) compared with those with S(2)/Te(2) co-occupancy model ($R_1$=0.0287, $wR_2$=0.0766), as well as having more isotropic thermal parameters. The Hamilton significance test[25] indicates that the final structural model, with the independent S(2) and Te(2) positions, is superior at greater than the 99.99% confidence level. The Bi-S(1) and Bi-S(2) bond lengths are 2.995(2) and 2.789(17) Å, respectively. This difference is consistent with the fact that the S atoms in sites Ch(1) and Ch(2) have coordination numbers to Bi of 6 and 3, respectively, and thus S(1) is more ionic than S(2).[24,26] The Bi-Te(2) bond length is 3.018(3) Å, which is close to that found in $Bi_2Te_2Se$.[27]

A schematic comparison of the crystal structures of $Bi_2Te_2Se$, $Bi_2TeSe_2$ and $Bi_2Te_{1.6}S_{1.4}$ is shown in the upper part of Figure 1. Similar to the $Bi_2Te_3$-$Bi_2Se_3$ system, where the Se atoms prefer to occupy the middle chalcogen layer in the quintuple-layer sandwich rather than randomly mixing with the Te atoms[24], the S atoms in $Bi_2Te_{1.6}S_{1.4}$ also prefer to occupy the middle chalcogen layer. In the two outer layers of the quintuple layer sandwich, the S and Te atoms mix with each other randomly in a 0.2:0.8 ratio, maintaining the space group symmetry R-3m commonly observed in the Tetradymite family. The lower part of Figure 1 shows a comparison of the $BiCh_6$ octahedra in $Bi_2Te_{1.6}S_{1.4}$ (left) and $Bi_2Te_3$ (right), as well as the aerial views of the corresponding $BiCh(2)_3$ layers. Compared with $Bi_2Te_3$ (a=4.395 Å), the *a* axis of $Bi_2Te_{1.6}S_{1.4}$ is reduced by about 5%, to 4.196 Å due to the smaller size of sulfur. This significantly strains the $BiCh_6$ octahedron in $Bi_2Te_{1.6}S_{1.4}$ due to the presence of the necessarily smaller Te(2)-Bi-Te(2) angle, and provides a rationalization for the nonstoichiometry of the phase: to help relieve this strain, the stable compositions incorporate some fraction of smaller S atoms in the Ch(2) layer. The fact that the Te(2) layer in $Bi_2Te_{1.6}S_{1.4}$ is further from the Bi layer than it is in $Bi_2Te_3$ (i.e. with an interplanar spacing of 1.80 Å compared with 1.72 Å) may reflect the structure's attempt to maintain a preferred Bi-Te(2) bond length within the strained lattice. Sb is smaller than Bi and thus the fact that lower S contents are required for stable compound formation in the Sb-substituted phase is consistent with this strain argument.

Powder X-ray diffraction patterns for the ground crystals of as-grown $Bi_2Te_{1.6}S_{1.4}$, annealed $Bi_2Te_{1.6}S_{1.4}$, and as-grown $Bi_{1.1}Sb_{0.9}Te_2S$ are shown in Figure 2. The peak positions can all be indexed using the rhombohedral unit cell of Tetradymite.[20] By comparing the patterns of as-grown and annealed $Bi_2Te_{1.6}S_{1.4}$, it can be seen that the quality of the $Bi_2Te_{1.6}S_{1.4}$ crystals deteriorates after annealing, resulting in broader peak shapes and poorer diffraction. The pattern of as-grown $Bi_{1.1}Sb_{0.9}Te_2S$ is as good as that of as-grown $Bi_2Te_{1.6}S_{1.4}$ at this level of structural characterization, indicating that it is a pure phase with good crystalline quality. A simple substitution of Sb for Bi in $Bi_2Te_{1.6}S_{1.4}$ without changing the Te/S ratio results in the presence of the impurity phase $Sb_2S_3$ in the crystal growth. Therefore lower S contents, for example with the formulas $Bi_{1.4}Sb_{0.6}Te_{1.8}S_{1.2}$ ($Bi_{1.4}Sb_{0.6}Te_{1.8}S_{1.2}$ grows large crystals, with only a few grains in samples ten inches long, via the Bridgeman-Stockbarger Method)[18] and $Bi_{1.1}Sb_{0.9}Te_2S$, were needed to grow the high quality crystals of partially Sb substituted materials.

Transport data is shown in Figure 3. High *n*-type carrier concentrations, around $5 \times 10^{19}$

cm$^{-3}$, were calculated from Hall coefficient measurements on single crystals of native Bi$_2$Te$_{1.6}$S$_{1.4}$. Bi$_2$Te$_{1.6}$S$_{1.4}$'s resistivity is relatively low, ranging from 6×10$^{-4}$ Ohm-cm (300K) to 3.5×10$^{-4}$ Ohm-cm (10K). A representative piece cut from a Bi$_{1.4}$Sb$_{0.6}$Te$_{1.8}$S$_{1.2}$ boule of volume several cm$^3$, shows a temperature independent resistivity, of around 10$^{-2}$ Ohm-cm and a temperature independent *n*-type carrier concentration of ~7×10$^{17}$/cm$^3$. Carrier concentrations for different parts of the boule differed by up to a factor of two, reflective of an Sb content that is not strictly homogeneous. Also shown in the figure is the characterization of a crystal taken from the boule of composition Bi$_{1.1}$Sb$_{0.9}$Te$_2$S. This crystal is *p*-type, and since it has a low *p*-type carrier concentration (1.6×10$^{16}$/cm$^3$) at 10 K, suggests that this Bi:Sb ratio is very close to the crossover between dominant *n*- and dominant *p*- type behavior for this phase.

The characterization by ARPES is shown in Figure 4, which includes the comparison of several Tetradymites. Figure 4(c) shows the bulk and surface band structures of Bi$_2$Te$_{1.6}$S$_{1.4}$, which displays a bulk band gap of about 200 meV at the Γ point. The single Dirac cone observed for its surface states shows that it is a 3D TI. Compared with Bi$_2$Te$_3$ (Fig. 4(a)), whose Dirac point is largely buried into the bulk valence bands,[11, 28] Bi$_2$Te$_{1.6}$S$_{1.4}$ has its Dirac point more fully exposed; its position with respect to the bulk valence band is similar to what is seen in Bi$_2$Te$_2$Se (Fig. 4(b)), which has a bulk band gap of about 300 meV at the Γ point. Detailed photon energy dependence ARPES measurements[29] indicate that the Dirac point is buried below the top of the valence band by approximately 50 meV in Bi$_2$Te$_2$Se, while similar analysis shows that it is above the bulk valence bands by approximately 30 meV in Bi$_2$Te$_{1.6}$S$_{1.4}$. Thus though the materials are similar both chemically and electronically, the relative energies of their bulk valence bands and Dirac points differ. Preliminary ARPES characterization of the Sb-substituted material Ba$_{1.1}$Sb$_{0.9}$Te$_2$S (Fig. 4(d)) shows that the chemical potential is now below the bulk conduction band, in the surface state energy regime, and further suggests that the Dirac point is more than 100 meV above the bulk valence band, making it substantially more exposed than is seen for non-Sb substituted Bi$_2$Te$_{1.6}$S$_{1.4}$ and Bi$_2$Te$_2$Se.

**Discussion and Conclusion**

Our crystal growth studies of Bi$_2$Te$_{2-x}$S$_{1+x}$ (0≤x≤1) show that within this composition regime the nominal composition $Bi_2Te_{1.5}S_{1.5}$ (x=0.5) melts nearly congruently, and grows monolithic single crystals of the γ Tetradymite phase with the composition $Bi_2Te_{1.6}S_{1.4}$. Similar to the $Bi_2Te_3$-$Bi_2Se_3$ system,[18, 30-33] the defect chemistry in $Bi_2Te_{2-x}S_{1+x}$ determines the dominant the carrier types and concentrations. The as-grown native material displays a large number of *n*-type carriers, with a typical concentration of ~ $10^{19}/cm^3$. Two major factors likely contribute to this result. Firstly, the dominant defect mechanism in pure $Bi_2Te_3$ is anti-site substitution, $Bi_2Te_3 \rightarrow 2Bi_{Te}' + 2h^\bullet + Te_2(g)$. With Bi bonding to the more electronegative S atoms in the case of $Bi_2Te_{1.6}S_{1.4}$, however, there will be much less flexibility to form anti-site defects,[31] decreasing the number of defects available that give rise to hole doping. Secondly, considering the higher volatility of S compared with Te or Se, S atoms have a stronger tendency to evaporate during the crystal growth, leaving electron-donating vacancies behind, i.e. $S_S \rightarrow V_S^{\bullet\bullet} + S(g) + 2e'$. $Bi_{1.1}Sb_{0.9}Te_2S$ shows the dominance of *p*-type carriers, following the general trend in Tetradymites that Sb-tellurides are favorable $Sb_{Te}'$ antisite defect formers, which increases the number of acceptors.

Finally, the ARPES measurements confirm that the sulfur-containing $Bi_2Te_{1.6}S_{1.4}$ Tetradymite is a 3D TI with a relatively isolated Dirac cone in its surface state dispersion, though similar to what is seen in $Bi_2Te_2Se$. Crystals at nearly the 1:1 Bi to Sb ratio, e.g. $Bi_{1.1}Sb_{0.9}Te_2S$, have the doubly advantageous characteristics of having low bulk carrier concentrations and a Dirac point that is further in energy from the top of the bulk valance band than for the pure Bi variant. First-principle calculations have suggested that $Bi_2Te_2Se$ and "$Bi_2Te_2S$," should be very similar to $Bi_2Te_3$, with indirect band gaps that increase from $Bi_2Te_3$, to $Bi_2Te_2Se$, to $Bi_2Te_2S$, consistent with the increasing electronegativity on going from Te, Se, to S. We note, however, that a smaller gap is observed at Γ in $Bi_2Te_{1.6}S_{1.4}$ compared to $Bi_2Te_2Se$; we speculate that this may be due to the presence of smaller spin orbit coupling in the current case, where the mass of the chalcogens is significantly lower. $Bi_2Te_{1.6}S_{1.4}$ offers another materials platform for studying the surface states on 3D TIs. More quantitative spectroscopic studies of the bulk and surface band structure of $Bi_2Te_{1.6}S_{1.4}$ would be of interest as would surface state transport studies performed on mixed $(Bi,Sb)_2(Te,S)_3$ crystals with more finely optimized compositions near $Bi_{1.1}Sb_{0.9}Te_2S$. Finally, STM (scanning tunneling microscopy) studies of the surface states in the presence of the randomly corrugated geometry of the outer S/Te layers of $Bi_2Te_{1.6}S_{1.4}$ may be of interest.

**Acknowledgements**

The crystal growth, crystal structure determination and electronic property characterization were supported by NSF grants DMR-0819860 and DMR-1005438 and AFOSR grant FA9550-10-1-0533. The ARPES characterization was supported by NSF grant DMR-0819860 and DOE grant DE-FG02-07ER46352.

## Table 1 – Structural Characterization of $Bi_2Te_{1.6}S_{1.4}$

(a) Refinement data

| | |
|---|---|
| Formula sum | $Bi_2Te_{1.6-\delta}S_{1.4+\delta}$ |
| | ($0 \leq \delta \leq 0.032$) |
| Space group | R-3m (No. 166) |
| a (Å) | 4.196(4) |
| c (Å) | 29.44(3) |
| V (Å3) | 448.89(70) |
| Z | 3 |
| Temperature (K) | 293(2) |
| F000 | 838 |
| Reflections | 173 |
| R1 (all reflections) | 0.0251 |
| R1 (Fo>4σ(Fo)) | 0.0247 |
| wR2 | 0.0700 |
| Rint/ R(σ) | 0.0412/0.0221 |
| GooF | 1.360 |

Position coordinates and thermal parameters
(b) Independent S(2)/Te(2) position model. Agreement factors: R1=0.0251, wR2=0.0700.

| Atom | Wyck. | x | y | z | Occ. | $U_{11}$ | $U_{33}$ | $U_{12}$ |
|---|---|---|---|---|---|---|---|---|
| Bi | 6c | 0 | 0 | 0.39316(2) | | 0.0100(3) | 0.0138(5) | 0.00501(17) |
| S(1) | 3a | 0 | 0 | 0 | | 0.0115(19) | 0.011(3) | 0.0057(10) |
| Te(2) | 6c | 0 | 0 | 0.21237(7) | 0.792(8) | 0.0105(6) | 0.0131(12) | 0.0053(3) |
| S(2) | 6c | 0 | 0 | 0.2266(12) | 0.208(8) | 0.0105(6) | 0.0131(12) | 0.0053(3) |

(c) S(2)/Te(2) Co-occupancy model. Agreement factors: R1=0.0287, wR2=0.0766.

| Atom | Wyck. | x | y | z | Occ. | $U_{11}$ | $U_{33}$ | $U_{12}$ |
|---|---|---|---|---|---|---|---|---|
| Bi | 6c | 0 | 0 | 0.39316(2) | | 0.0100(4) | 0.0142(5) | 0.00499(19) |
| S(1) | 3a | 0 | 0 | 0 | | 0.011(2) | 0.011(3) | 0.0056(11) |
| Te(2) | 6c | 0 | 0 | 0.21278(6) | 0.790(10) | 0.0108(7) | 0.0203(10) | 0.0054(3) |
| S(2) | 6c | 0 | 0 | 0.21278(6) | 0.210(10) | 0.0108(7) | 0.0203(10) | 0.0054(3) |

**Figures**

**Fig. 1 (Color on line).** (a) Comparison of quintuple-layers of $Bi_2Te_2Se$, $Bi_2Se_2Te$, and $Bi_2Te_{1.6}S_{1.4}$, atomic positions with colored half hemispheres are occupied by mixtures of chalcogens. (b) Side views of the $BiCh_6$ octahedra in $Bi_2Te_{1.6}S_{1.4}$ (left) and $Bi_2Te_3$ (right), as well as the aerial views of the corresponding $BiCh(2)_3$ layers. The interplanar spacings and the lengths of *a* axes, i.e. the in-plane chalcogen-chalcogen separations, are labeled.

**Fig. 2 (Color on line).** X-ray powder diffraction patterns for as-grown $Bi_2Te_{1.6}S_{1.4}$ (lower, black pattern), annealed $Bi_2Te_{1.6}S_{1.4}$ (middle, red pattern), and as-grown $Bi_{1.1}Sb_{0.9}Te_2S$ (upper, blue pattern). The expected peak positions for the Tetradymite structure for Cu Kα radiation are shown.

**Fig. 3 (Color on line).** The Temperature dependent basal plane resistivities for single crystals of native $Bi_2Te_{1.6}S_{1.4}$, $Bi_{1.4}Sb_{0.6}Te_{1.8}S_{1.2}$, and $Bi_{1.1}Sb_{0.9}Te_2S$. The carrier concentrations calculated from Hall measurements are listed in the figure. The inset shows the temperature dependent net *p*-type carrier concentration determined from Hall measurements on the crystal of $Bi_{1.1}Sb_{0.9}Te_2S$.

**Fig. 4 (Color on line).** Comparison of the ARPES characterizations of the band structures of (a) $Bi_2Te_3$, (b) $Bi_2Te_2Se$, (c) $Bi_2Te_{1.6}S_{1.4}$ and (d) $Bi_{1.1}Sb_{0.9}Te_2S$ along the M-Γ-M high symmetry momentum directions. The features are identified in panel (c): BVB and BCB denote the Bulk Valence Band and Bulk Conduction Band, respectively. The sharp V-shape dispersions are from the surface states (SS), with the nodes of the Dirac cones denoted as the Dirac Points. The data in (a-c) are taken with 10 eV photons, and the lower resolution data in (d) are taken with 60 eV photons.

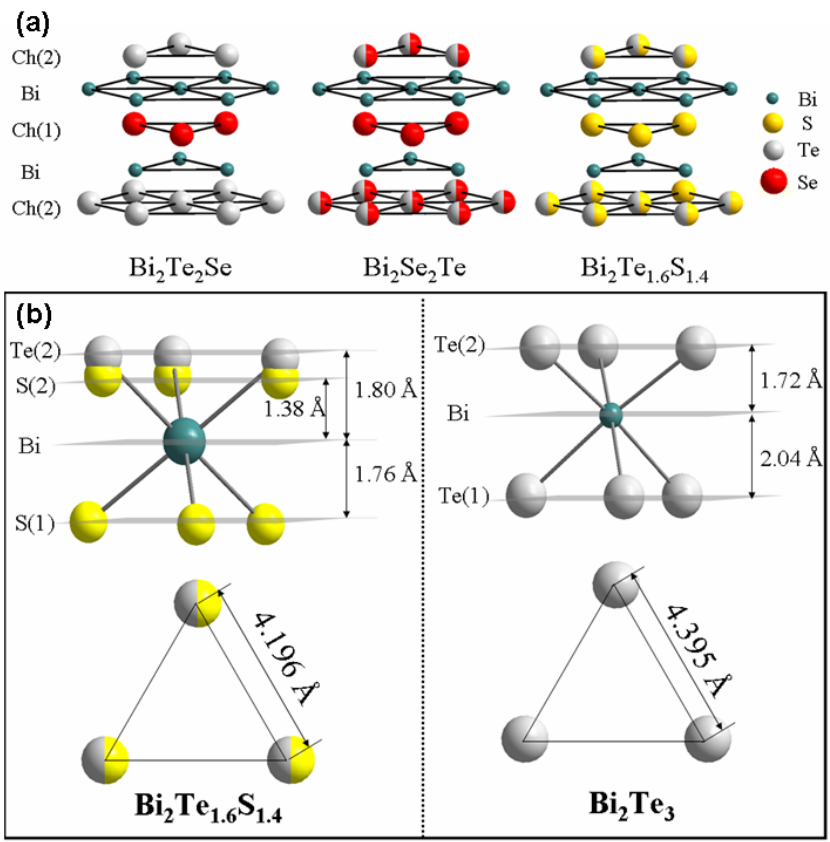

**Fig. 1**

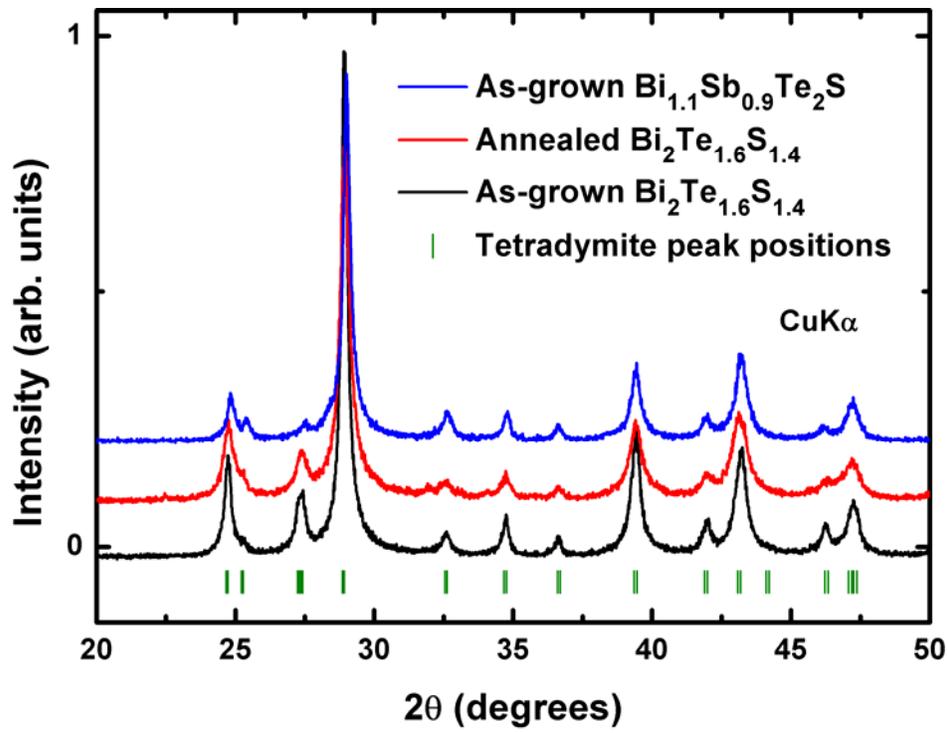

Fig 2

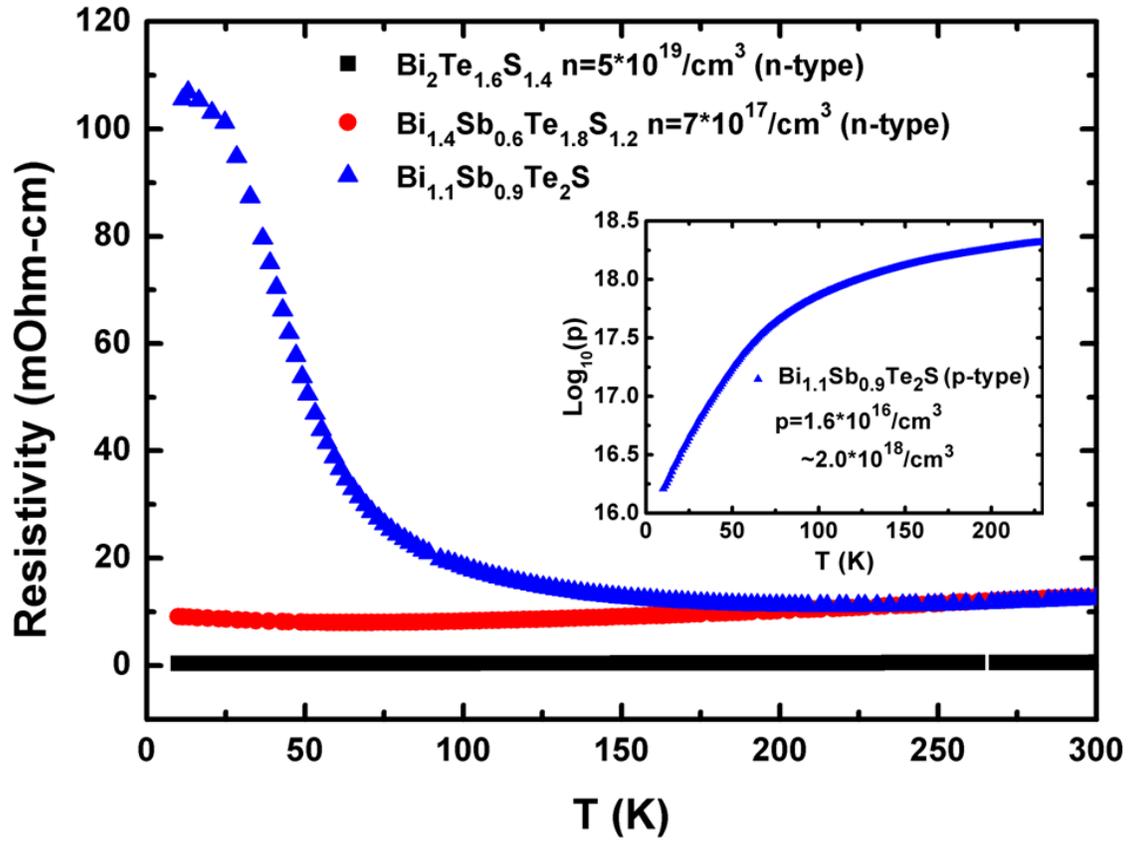

**Fig. 3**

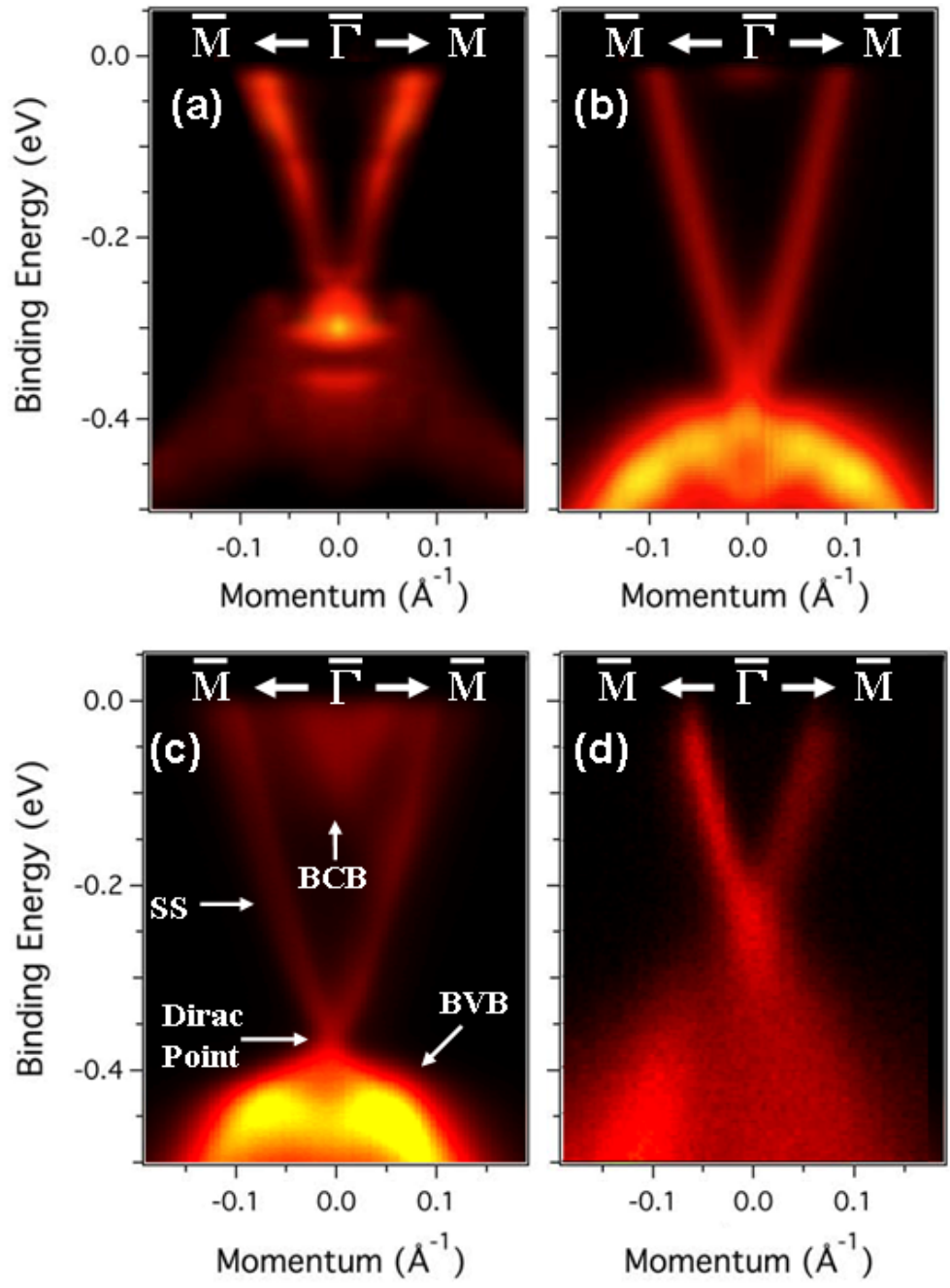

Fig 4